\documentclass[prd,nofootinbib,preprint,superscriptaddress,letter]{revtex4}
\pdfoutput=1

\usepackage{float}
\usepackage{comment}
\usepackage{graphicx}
\usepackage{physics,slashed}
\usepackage{amsfonts,amsmath,amssymb}
\usepackage{mathrsfs}
\usepackage{bm}

\usepackage{epsfig}
\usepackage{graphicx}   
\usepackage{subfigure}  
\usepackage{verbatim}   
\usepackage{color}      
\usepackage{url}
\definecolor{nicecol1}{rgb}{0.56,0.,1.}
\definecolor{nicecol2}{rgb}{1.,0.3,0.8}
\definecolor{nicecol3}{rgb}{0.,1.,0.}
\usepackage{array}
\usepackage{dcolumn}
\usepackage{multirow}
\usepackage{cprotect}
\usepackage{framed}
\usepackage{setspace}
\usepackage{enumitem}
\usepackage{graphicx}
\setlist{nolistsep}

\newcommand{\GeV}{\rm GeV}

\newcommand{\qcd}{$\Lambda_{QCD}$}

\renewcommand\[{\begin{equation}}
\renewcommand\]{\end{equation}}

\begin{document}

\title{\vspace{-20mm}
\begin{flushright}
\normalfont{\normalsize{DIAS-STP-24-01}}\\
\normalfont{\normalsize{UCI-HEP-TR-2024-03}}
\end{flushright}
\hfill ~\\[-5mm]
\textbf{\Large
Constraints on Variation of the Weak Scale\\ from Big Bang Nucleosynthesis}}

\author{Anne-Katherine Burns}
\email{annekatb@uci.edu}
\affiliation{Department of Physics and Astronomy, University of California, Irvine, CA 92697 USA}
\author{Venus Keus}
\email{venus@stp.dias.ie}
\affiliation{Dublin Institute for Advanced Studies, School of Theoretical Physics, 10 Burlington road, Dublin, D04 C932, Ireland}
\author{Marc Sher}
\email{mtsher@wm.edu}
\affiliation{High Energy Theory Group, Physics Department, William \& Mary, Williamsburg, VA 23187, USA}
\author{Tim M.P. Tait}
\email{ttait@uci.edu}
\affiliation{Department of Physics and Astronomy, University of California, Irvine, CA 92697 USA}

\vspace{4mm}

\begin{abstract}
\noindent
    Recently, the EMPRESS collaboration has included new data in the extraction of the primordial $^4$He abundance from Big Bang Nucleosynthesis (BBN), resulting in a determination that differs from the previous value and from theoretical expectations.  
   There have been several studies attempting to explain this anomaly which involve variation of fundamental constants between the time of BBN and the present.  Since the Higgs vacuum expectation value (vev) is the only dimensionful parameter in the Standard Model and it is already known to vary during the electroweak phase transition, we consider the possibility that the vev is slightly different during BBN compared to its present value.
     A modification of the vev changes not only particle masses but also affects, through mass thresholds, the QCD confinement scale.  We use the recently developed PRyMordial program to study this variation and its impact on the $^4$He and deuterium abundances.  We find that bounds on $|{\delta v}/{v}|$ are approximately $0.01$, and that the EMPRESS result can be explained within $2\sigma$ if  $0.008 < {\delta v}/{v}< 0.02$, but at the cost of worsening the current $2\sigma$ discrepancy in the deuterium abundance to over $3\sigma$.
  
\end{abstract}

\vspace{4mm}

\date{\today}

\maketitle

\section{Introduction}
The $\Lambda$CDM model has been extraordinarily successful in describing the cosmological history and evolution of the Universe.   However, there are some anomalies in the model.   The most well-known is the Hubble tension, in which the value of the Hubble parameter measured from the cosmic microwave background (CMB) by Planck~\cite{Planck:2018vyg}, $H_0 = 67.4 \pm 0.5\ {\rm km/s/Mpc}$, differs by $5 \sigma$ from the measurement of Cepheids and Type-Ia supernovae by the SH0ES collaboration\cite{Riess:2021jrx}, $H_0 = 73.0 \pm 1.0\ {\rm km/s/Mpc}$.   There have been numerous studies trying to explain this discrepancy (see Ref.~\cite{Dainotti:2023yrk} for an extensive review).    Another anomaly concerns the ${}^4$He abundance, $Y_p$, from Big Bang Nucleosynthesis (BBN).    A recent report by EMPRESS~\cite{Matsumoto:2022tlr} adds measurements from 10 additional extremely metal-poor galaxies which, combined with the previously existing dataset, results in a fit to $Y_p = 0.2370 \pm 0.0033$ which differs from the value obtained based only the pre-existing data~\cite{Kurichin:2021ppm,Hsyu:2020uqb,Aver:2021rwi}.    This has led to several studies, such as the implications of the result for the lepton asymmetry~\cite{Kawasaki:2022hvx,Burns:2022hkq,Borah:2022uos,Escudero:2022okz}.

The variation of fundamental constants is a subject that has been around since Dirac's large number hypothesis~\cite{dirac}.   Most work in the past has focused on variation in the fine-structure constant~\cite{Bergstrom:1999wm,Nollett:2002da,Mosquera:2007vr}
and the electron mass~\cite{Barrow:2005qf,Scoccola:2008yf} and both have been used to study the above anomalies.   Variation of the fine-structure constant to solve the Hubble tension was proposed in a mirror dark sector model recently~\cite{Zhang:2022ujw}.  Variation of the electron mass~\cite{Barrow:2005qf} has also been proposed as a solution to the Hubble tension in Ref.~\cite{Seto:2022xgx} in which they also studied the effects on Big Bang Nucleosynthesis; a mechanism for such a variation was proposed in Ref.~\cite{Solomon:2022qqf}.   A detailed series of papers studying variation of both the fine-structure constant and the electron mass is by Hart and Chluba~\cite{Hart:2017ndk,Hart:2019dxi,Hart:2021kad}.
The helium anomaly is more recent.  It was shown by Seto, Takahashi and Toda~\cite{Seto:2023yal} that a variation in the fine-structure constant alone would be sufficient to explain the $Y_p$ anomaly.

While one can consider variation of the fine-structure constant and the electron mass, we feel that it would be more reasonable to consider variation in the vacuum expectation value (vev) of the Higgs field (or, equivalently, the Higgs mass term).   The vev is the only dimensionful parameter in the Standard Model, and it is also the most mysterious, in the sense that its value is many, many orders of magnitude smaller that one would naturally expect.   In addition, the Standard Model vev is cosmologically dynamical, and is predicted to vary substantially during the electroweak phase transition.  At high temperature, the gauge symmetry is restored \cite{Dolan:1973qd,Weinberg:1974hy} and the vev is zero.  As the universe cools, a transition occurs and the vev increases to the low temperature value observed today.   
Thus, it would seem to be more reasonable to consider variations in the vev rather than in one of the many dimensionless parameters.   For example, one can imagine an extended model in which the Higgs vev slowly rolls to its current value but has not quite reached it at the MeV temperature scale. One might wonder if that might affect later observables such as the CMB, but that occurs at the $\sim$eV temperature scale and we presume that the slow roll would have essentially completed by then.

The possible effects of a changing vev on BBN was first discussed in 1988 by Dixit and Sher~\cite{Dixit:1987at}, but this was a crude calculation that only considered the ${}^4$He abundance.   There have since been several studies~\cite{Yoo:2002vw,Chamoun:2005xr,Gassner:2006fr,Landau:2008re} of BBN due to a changing vev and other studies~\cite{Muller:2004gu,Coc:2006sx} on the effects of changing many parameters, including the vev.   These papers all include the effects not only on the electron mass but also on quark masses, which subsequently affect the pion and other meson masses and thus the strong force (leading, for example, to changes in the deuteron binding energy).      Other works did not directly discuss modifying the vev, but only $\Lambda_{QCD}$~\cite{Kneller:2003xf}, the deuteron binding energy~\cite{Dmitriev:2003qq} and quark masses~\cite{Bedaque:2010hr}.  Finally, there has also been work done looking at the effect of changing the strength of the weak interaction at late times \cite{Ferrero:2010ab}. Like these works, our approach is phenomenological and we will not present a specific model with a changing vev at the time of BBN.    A recent paper~\cite{Laverda:2024qjt} did present a model in which the Higgs field may not achieve a thermal spectrum at the time of nucleosynthesis - the effects on the vev are unclear.    Further investigation of this model would be interesting.

However, these works did not include the effects of a changing Higgs vev on the value of $\Lambda_{QCD}$, which clearly will affect strong interaction dynamics.  As noted by Agrawal, \textit{et al.}~\cite{Agrawal:1997gf,Agrawal:1998xa}, a variation in the Higgs vev will modify all of the quark masses and this in turn will impact the running of the strong coupling constant through the quark thresholds.   The effect is somewhat smaller (they find $\Lambda_{QCD}$ varying by roughly $(v/v_0)^{0.25})$), but can still be substantial.   These effects were included in Ref.~\cite{Jeltema:1999na} which was an anthropic principle based study of the triple alpha process in stars, but have not been included in analyses of BBN.

Since these above works, we have learned much more about the current baryon-photon ratio, the nuclear abundances and have better codes~\cite{Burns:2023sgx}. One such code, PRyMordial was created to fill several gaps that currently exist in the landscape of publicly available BBN codes. Namely, the code allows for easy and flexible exploration of a variety of new physics models, including the investigation of varying input parameters on both $N_{eff}$ and the final abundance values. PRyMordial also allows users to examine the uncertainties in different sets of thermonuclear reaction rates and to scale these chosen reaction rates as desired. In addition, PRyMordial calculates the thermodynamics of the plasma from first principles, the results of which serve as the initial conditions for the full calculation of neutron to proton conversion. Unlike other codes, PRyMordial is written in \texttt{Python}, allowing its users to perform Monte Carlo (MC) analyses with ease by interfacing with standard MC libraries. In order to maximize efficiency, the code also has an option to use \texttt{Julia} to speed up the computation. PRyMordial has already been used by several groups to study the effects of a variety of new physics scenarios on BBN~\cite{Burns:2022hkq,Meissner:2023voo,Schoneberg:2024ifp,Giri:2023pyy,Hong:2023fcy,Bianchini:2023ubu,Chowdhury:2022ahn}. In this paper, using PRyMordial, we will calculate the abundances of light nuclei as a function of the Higgs vev including all of these effects and in light of the EMPRESS result.

\section{BBN as a Function of the Higgs vev}

In looking at the effect of changing the Higgs vev on the production of light elements, there are several important parameters that determine the abundance of these light elements: the neutron-proton mass difference, the deuteron binding energy, and the rates of several key thermonuclear reactions involving light nuclei.
In this section, we consider how each of these vary with the Higgs vev, in turn. 

\subsection{Neutron-Proton Mass Difference}

In order to calculate the way in which the neutron-proton mass difference changes with a changing vev, we first considered the change in the masses of their constituent quarks. When the vev is varied, the quark masses receive fractional corrections of ${\delta v}/{v}$.   While the bulk of the neutron and proton masses are the result of confinement, their masses, and most importantly their mass difference, is sensitive to the change in quark masses.  As is well known, the down quark is roughly $2.5$ MeV heavier than the up quark, but the neutron is only $1.293$ MeV heavier than the proton.   The discrepancy is due to the electromagnetic energy of the up quarks within the proton, which is thus approximately $1.2$ MeV, independent of the Higgs vev.   As a result, we take the neutron-proton mass difference in MeV to be
\begin{equation}
\frac{m_n - m_p}{\rm MeV} = 2.493 ~(1+\frac{\delta v}{v}) - 1.2\label{npmass} ~.
\end{equation}
This mass difference has a direct impact on the rate of the neutron-proton inter-conversion which occurs via six reactions which, along with protons and neutrons, also involves electrons and electron neutrinos. At temperatures above about 1 MeV, these reactions were in chemical equilibrium and conversion happened freely and regularly. In his 1972 book, Weinberg~\cite{Weinberg:1972kfs} lists the reaction rates for these six processes and one can see that both the variation in the neutron-proton mass difference and the electron mass, which varies as ${\delta v}/{v}$ like the quarks, will suffice to determine the variation in these processes. By including information about the neutron lifetime, the final ratio of the number of neutrons to the number of protons at the beginning of BBN can be determined. Because almost all of the neutrons end up in $^4$He, this ratio allows us to crudely calculate the $^4$He abundance. However, to more precisely calculate the  $^4$He abundance, along with the abundances of deuterium, tritium and $^3$He nuclei, it is necessary to include in the calculation at least twelve thermonuclear reactions, all of which play essential roles in the formation of nuclei during BBN.

\subsection{$n \leftrightarrow p$ Conversion Rates}

\begin{figure}[th!]
    \begin{minipage}{0.48\textwidth}
        \centering
        \includegraphics[width=\linewidth]{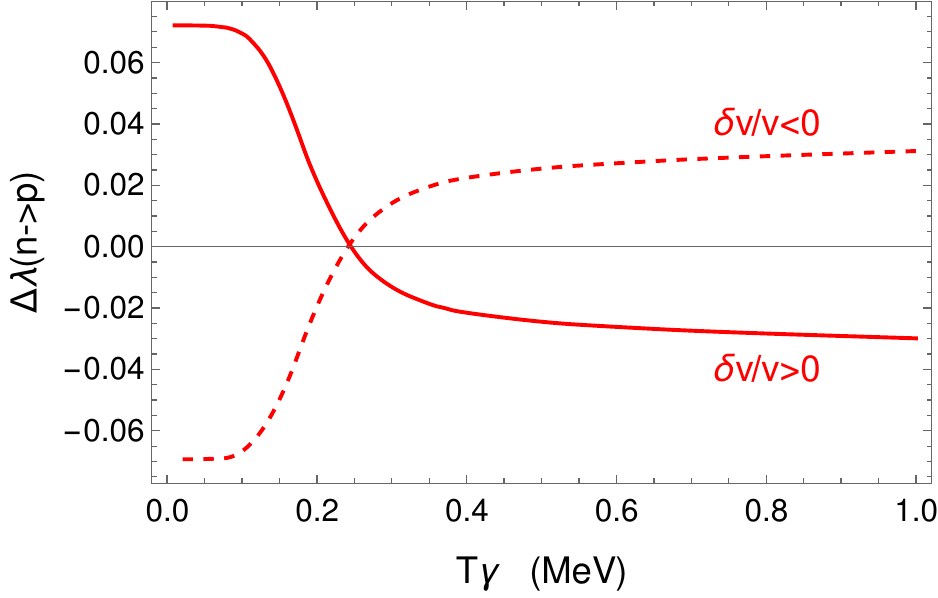}
        \label{fig:ntop}
    \end{minipage}\hfill
    \begin{minipage}{0.48\textwidth}
        \centering
        \includegraphics[width=\linewidth]{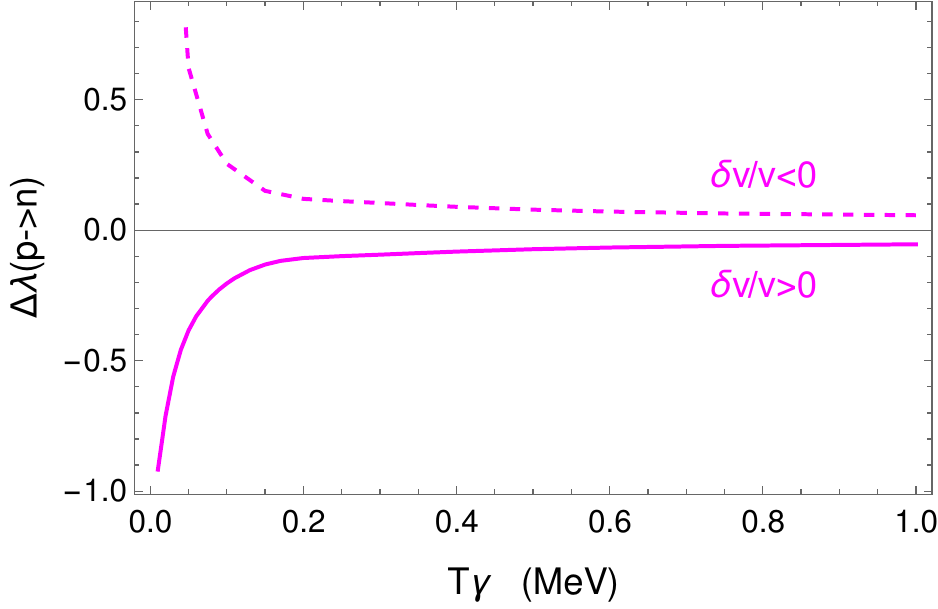}
        \label{fig:pton}
    \end{minipage}
    \caption{The fractional change in the $\lambda(n \to p)$ and $\lambda(p \to n)$ rates for $d\equiv |{\delta v}/{v}| = 0.01$. Note that the fractional change in $\lambda(p \to n)$ gets very large at low temperatures since the rate itself vanishes at low temperature. For each rate, we include a factor of $1/(1+d)^4$ coming from the shift in the $W$ mass.}
    \label{fig:pton_ntop}
\end{figure}

The theoretical expressions for the total $n\to p$ and $p\to n$ conversion rates are given in Weinberg~\cite{Weinberg:1972kfs}.    Using Eq.~\eqref{npmass} for the neutron-proton mass difference and scaling the electron mass fractionally as ${\delta v}/{v}$, the relationship between the electron and neutrino temperatures is slightly modified.
Putting these effects together with the factor of $1/(1+ {\delta v}/{v} )^4$ coming from the shift in the $W$ boson mass,
the resulting fractional change for the $n \to p$ and $p \to n$ conversion rates as a function of $T_\gamma$, the radiation temperature, for $d \equiv |{\delta v}/{v}| = 0.01$, are shown in Figure 1.  The fractional changes in the rates scale approximately linearly with $\delta v$.

\subsection{Deuteron Binding Energy}

In addition to these thermonuclear reaction rates, the binding energy of the deuteron is an important component of the calculation. Its relatively low value leads to the well-known deuterium bottleneck, which refers to the time period during which nucleosynthesis had begun, but the average temperature of photons in the bath was higher than the deuterium binding energy. As a result, almost immediately after deuterium formed in the aforementioned process, it would photo-disassociate, and until the photon bath had cooled below this value, BBN was unable to proceed. 

In order to understand the way in which the deuteron binding energy varies with a changing vev, we first compute the way in which the pion mass varies.   The pion mass in QCD is given by $m_\pi^2 \simeq (m_u+m_d) f_\pi$.   While the quark masses scale linearly with $\delta v$, the value of $f_\pi$ is proportional to $\Lambda_{QCD}$.    As noted earlier, this will scale differently from the quark masses and will depend on the vev through mass thresholds.    In Appendix A, we show that $\Lambda_{QCD}$ scales as $(\delta v)^{0.25}$, and thus, the pion mass scales as $(\delta v)^{1.25/2}$.

We now turn to the deuteron binding energy itself.   Since we are interested in the impact of a small change in the vev, high precision in the standard calculation is not necessary.   We model the nucleon-nucleon potential for the deuteron as an exchange of pion, $\sigma$ and $\omega$ mesons as discussed in the review article by Meissner~\cite{Meissner:1987ge}.   The pion mass, as discussed above, scales as $\left({\delta v}/{v}\right)^{1.25/2}$.    The $\omega$ mass comes primarily from QCD and so scales as $\left({\delta v}/{v}\right)^{0.25}$.    The $\sigma$ is believed to be a two-pion correlated state.   Lin and Serot~\cite{Lin:1990cx} calculated the $\sigma$ mass in terms of the pion-nucleon coupling, the pion mass and the nucleon mass.   Varying the masses in their expressions, we find that the $\sigma$ mass is insensitive to the pion mass, scaling as \qcd.     

\begin{figure}[h!]
    \includegraphics[width=0.85\linewidth, angle=0, clip]{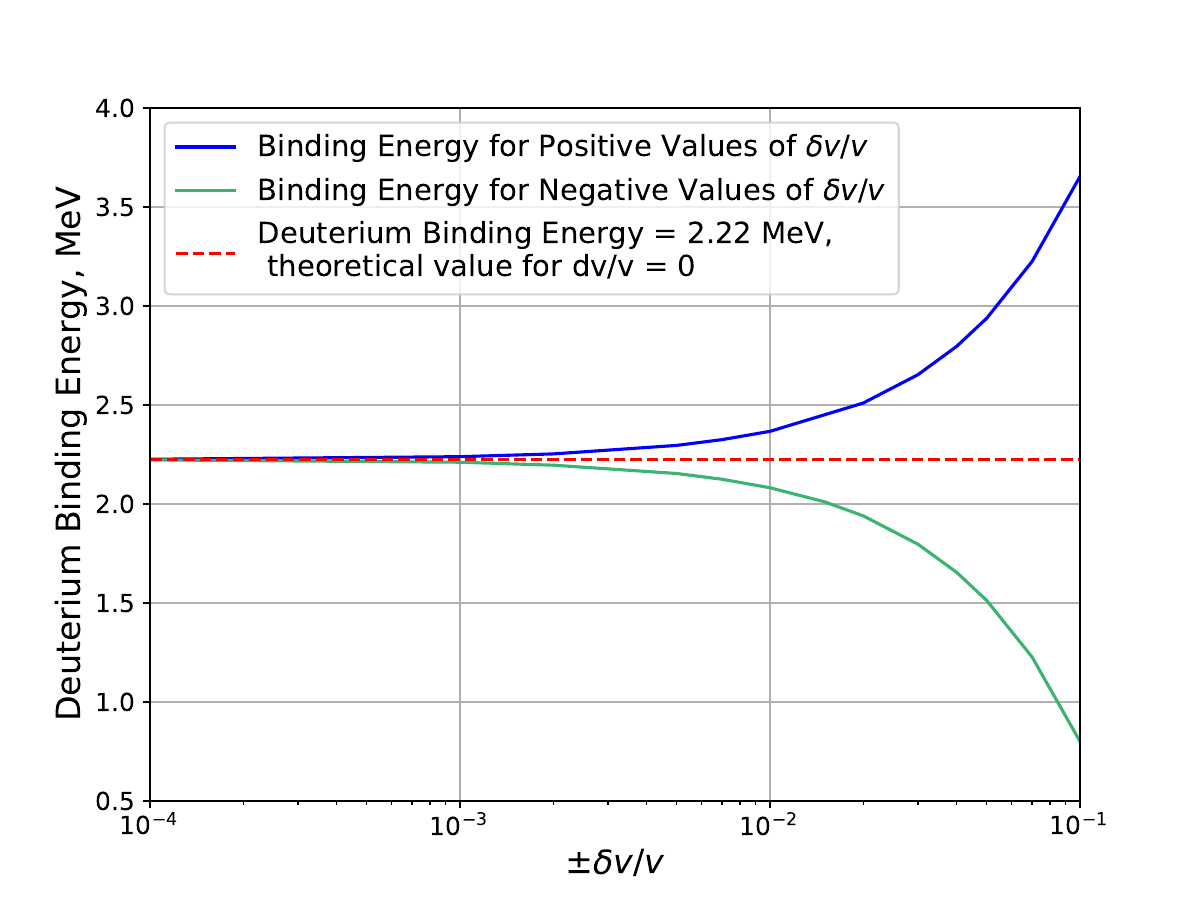}
    \caption{Value of the deuterium binding energy with varied Higgs vev for positive and negative values of ${\delta v}/{v}$.}
    \label{fig:dbechangeneg}
\end{figure}

The net result of $\pi$, $\omega$, and $\sigma$ exchange results in a potential energy function for the binding of a neutron and a proton which can be modelled as the sum of three Yukawa potentials - repulsive from the $\omega$ and attractive from $\sigma$ and $\pi$.
Each potential has a corresponding coupling constant indicated by $g_\pi$, $g_\sigma$ and $g_\omega$.  These couplings are in principle determined by QCD, but cannot be computed in perturbation theory, and so we constrain them by requiring that the resulting binding energy of the deuteron matches its experimentally determined value of $2.2$ MeV at $d \equiv |{\delta v}/{v}|=0$.  This selects families of viable parameters which can be found for $d=0$ by adjusting these three parameters such that the solution to the Schr\"{o}dinger equation gives a binding energy of $2.2$ MeV.   One such solution\footnote{We have examined other values of $g_\pi$, $g_\sigma$ and $g_\omega$ which give the correct $d=0$ binding energy of $2.2$ MeV, and find that they make very similar predictions for small $d$ to the ones we have chosen.}, for example, has $g_\pi = 11.97$, $g_\sigma = 8.46$, $g_\omega = 21.19$.   These are dimensionless parameters and are not expected to change much in response to small changes in the vev.  Results for small $d \neq 0$ are obtained by adjusting the meson masses for the particular $d$ of interest as described above, and recalculating the binding energy.   The resulting binding energies are presented in Figure~\ref{fig:dbechangeneg} for positive and negative values of ${\delta v}/{v}$.   Values of $|{\delta v}/{v}| > 0.1$  lead to either very large binding energies or an unbounded deuteron for positive and negative values, respectively. Since both very large binding energies and an unbounded deuteron, lead to predictions for the light elements abundances very far outside of what has been observed, we find it unnecessary to consider values of ${\delta v}/{v}$ outside of this range.

\subsection{Thermonuclear Reaction Rates}

With the change in the $n\leftrightarrow p$ processes and the deuteron binding energy in hand, we turn to the other nuclear processes in BBN.   There are 12 key reactions which are needed to predict deuterium and helium abundances and an additional 51 reactions to accurately predict the lithium abundance which involve the binding energies and matrix elements of heavier nuclei.

These reaction rates are typically determined empirically, making it difficult to predict their dependence on the vev from first principles.   However, they do not involve the weak interactions and are not as sensitive to the pion mass and thus on dimensional grounds, we assume their rates vary as $\Lambda_{QCD}$, or $v^{0.25}$, which is expected to be a decent approximation since the pion interactions are long-range.    

\section{Primordial Abdundances}

Assembling all of the changes in the inputs to BBN described above, we implement them in PRyMordial to determine the abundances of primordial helium and deuterium as a function of ${\delta v}/{v}$.  Starting from a calculated set of initial conditions, including the ratio of photon to neutrino temperatures, the light element abundances are determined via a network of Boltzmann
equations. This system of equations is solved in three steps. First, the neutron to proton ratio at the temperature of neutron freeze out is determined by analyzing the neutron to proton conversion rates.\footnote{For explicit formulas, please see equations 15.7.14 and 15.7.15 in Weinberg's book \textit{Gravitation and Cosmology: Principles and Applications of the General Theory of Relativity} \cite{Weinberg:1972kfs}.} Next, using the calculated abundance values of protons and neutrons, the network of Boltzmann equations using 18 of the 63 thermonuclear rates is evolved down to the temperature of deuterium photo-disassociation. Finally, the network is further evolved down to temperatures of $\mathcal{O}$(keV) to determine the final primordial abundance values of each light element.

The result for $Y_p$, which characterizes the abundance of $^4$He, is shown in Figure 3 for both positive and negative values of ${\delta v}/{v}$.
\begin{figure}[ht!]
    \includegraphics[width=0.85\textwidth,angle=0,clip]{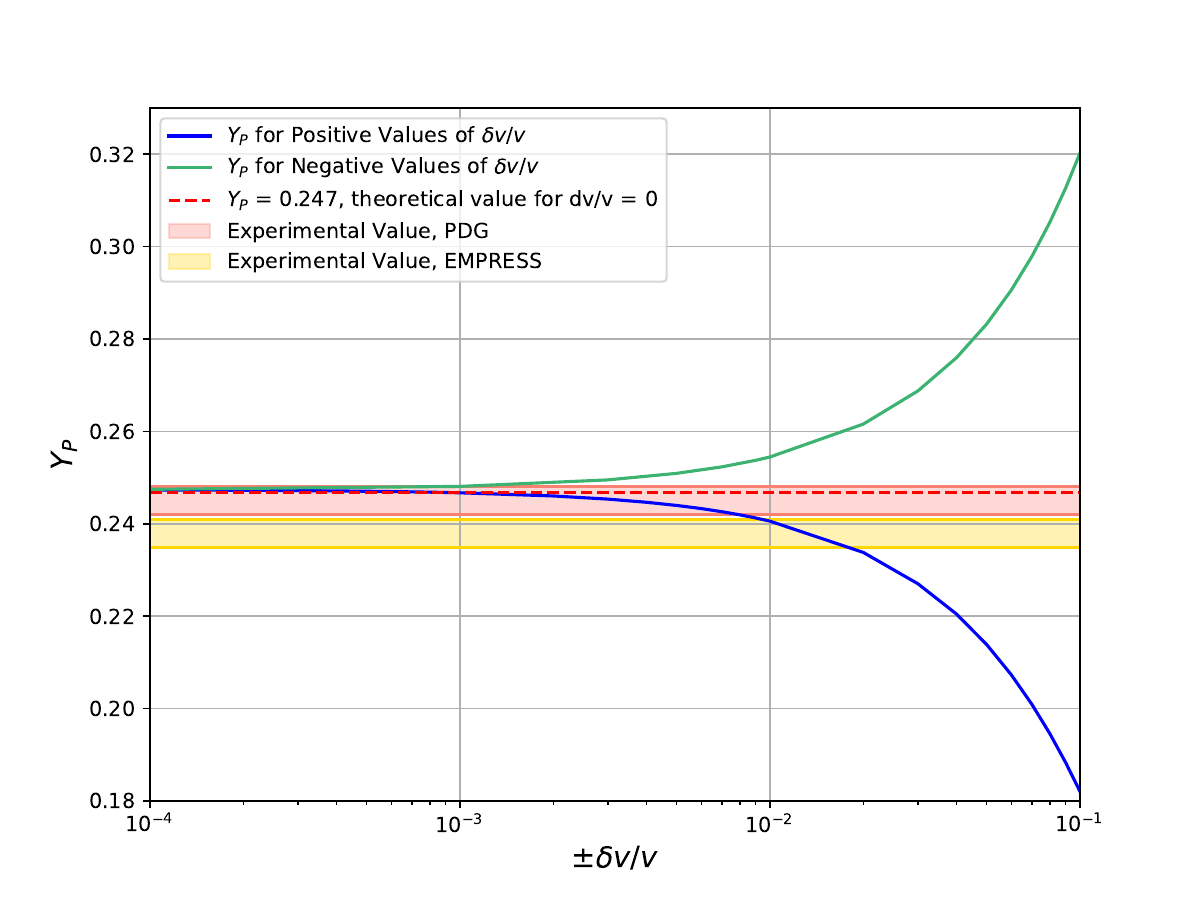}
\caption{Value of the abundance of $^4$He, $Y_p$, with varying Higgs vev for positive and negative values of ${\delta v}/{v}$.  The red dashed line is the standard result without a change in the Higgs vev and the blue and green lines are the results with the change.  The coral and gold boxes give the experimental values from the Particle Data Group and from the recent EMPRESS experiment.}
    \label{fig:hebechangeneg}
\end{figure}
The blue and green lines correspond to the predicted $^4$He abundance from PRyMordial for positive and negative values of ${\delta v}/{v}$, respectively.   The uncertainty for $Y_p$ is negligible.  The experimental value from Refs.~\cite{Kurichin:2021ppm,Hsyu:2020uqb} (as listed in the Particle Data Group~\cite{Workman:2022ynf} summary) and the $2\sigma$ uncertainties are shown in shaded pink, and the recent result from EMPRESS~\cite{Matsumoto:2022tlr} in shaded yellow.    For negative ${\delta v}/{v}$, the $2\sigma$ upper bound on the magnitude is approximately $0.001$, whereas for positive ${\delta v}/{v}$, the magnitude must be less than $0.008$ based on the PDG result and less than $0.02$ using the EMPRESS result.   If the EMPRESS result turns out to be correct, $\delta v = 0$ would be mildly excluded, and $\delta v$ between $0.008$ and $0.02$ would be able to explain the $^4$He abundance.

\begin{figure}[tb]
\includegraphics[width=0.85\textwidth,angle=0,clip]{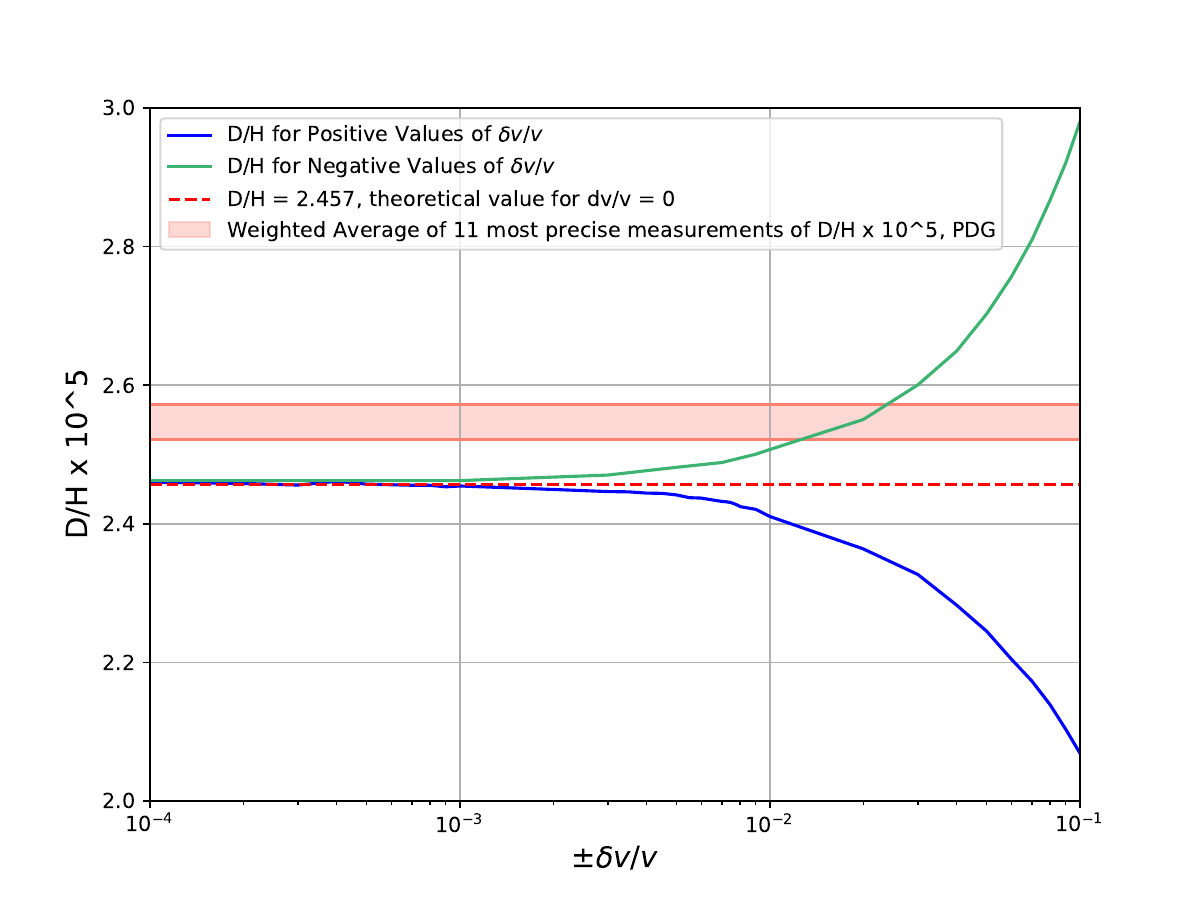}
\caption{Value of the abundance of deuterium with varying Higgs vev for positive and negative values of ${\delta v}/{v}$.  The red dashed line is the standard result without a change in the Higgs vev and the blue and green lines are the results with the change.  The coral box gives the experimental values from the Particle Data Group.  The error in the theoretical curves are similar to the error in the experimental values, as discussed in the text.}
\label{fig:dchangeneg}
\end{figure}

In Figure 4, we show the results for the deuterium abundance.   Here, the theoretical calculation of the abundance has an uncertainty of $0.05$, which is similar to the error in the experimental value \cite{Workman:2022ynf,Kislitsyn:2024jvk}.  Thus, the $\delta v = 0$ limit has roughly a $2\sigma$ discrepancy between the theoretical and experimental values.   For negative values of ${\delta v}/{v}$, we see that the magnitude of ${\delta v}/{v}$ must be less than approximately $0.025$, but for positive ${\delta v}/{v}$ the magnitude must be substantially smaller.   This is somewhat unfortunate since positive ${\delta v}/{v}$ is needed to explain the EMPRESS result for $Y_p$, but that would make the discrepancy with deuterium much worse.   Thus changing the vev can explain one, but not both, of the anomalies.

\section{Conclusions}

Originally motivated by anthropic arguments, many have studied the possibility that some of the constants of nature are time-dependent.   Astrophysical studies have bounded some parameters as have some cosmological studies.   Recently, the EMPRESS collaboration reported a measurement of the primordial $^4$He abundance which is somewhat over $3\sigma$ from the theoretical value.  There is also a small $2\sigma$ discrepancy in the deuterium abundance.   In this paper, we have considered the possibility that a difference in the Higgs vev between the time of BBN and the present could be responsible for one or both of these discrepancies.  We choose the Higgs vev to vary because it is the only dimensionful parameter of the Standard Model and is already known to vary during the earlier electroweak phase transition.

This was studied some time ago, but previous authors did not include the fact that the QCD scale will also vary if the quark masses vary, due to threshold effects.  They also did not use the latest BBN codes, which are more accurate than previous ones.   The varying quark masses and QCD scale will have a substantial effect on all nuclear reaction rates as well as the binding energy of the deuteron.

We find that the $^4$He abundance can be noticeably affected if the change in the Higgs vev is a few parts per thousand or more.   If ${\delta v}/{v}$ is between $0.008$ and $0.02$, then the prediction fits well within the EMPRESS calculation error bars.  If one instead uses the older PDG results, then one must (at $2\sigma$) have ${\delta v}/{v}$ less than $0.008$ and greater than $-0.001$.

The deuterium abundance is also affected. Here, a positive value of ${\delta v}/{v}$ makes the discrepancy with the theoretical prediction worse.   A negative value will fit within $2\sigma$ as long as its magnitude is less then $0.025$.     Together, we see that no value will be able to explain both the EMPRESS and deuterium anomalies, but can certainly explain either one.   Clearly, more experimental results for the $^4$He abundance are needed as are more theoretical studies of the deuterium abundance. 

\acknowledgments

We are very grateful for conversations with Meyer and Mei\ss{}ner related to their Ref. \cite{Meyer:2024auq}, which revealed a coding error in our earlier results but confirmed our over-all conclusions.

We thank Albert Stadler for discussions.
AKB and TMPT thank Mauro Valli for previous collaboration on PRyMordial.
The work of TMPT is supported in part by the U.S.\ National Science Foundation under Grant PHY-2210283.   The work of MS is supported by the U.S.\ National Science Foundation under Grant PHY-2112460. The work of AKB was supported in part by Grant No. NSF PHY-1748958 to the Kavli Institute for Theoretical Physics (KITP), the Heising-Simons Foundation, and the Simons Foundation (216179, LB). VK acknowledges financial support from the Science Foundation Ireland Grant 21/PATHS/9475 (MOREHIGGS) under the SFI-IRC Pathway Programme.

\appendix
\section{The evolution of the strong coupling constant}\label{Alpha-s}

A crude approximation to the scaling is obtained by integrating the one-loop renormalization group equations for the strong coupling constant from a scale well above the top quark mass down to somewhat below the charm quark mass and then using the standard formula to deduce \qcd.   Using a mass-independent renormalization scheme at one-loop, for $Q$ much larger than $2m_{t}$, one has
$$
\frac{1}{\alpha(1\ \GeV)^2} - \frac{1}{\alpha(Q^2)} = \frac{1}{12\pi}(21 \log 4m_{t}/Q^2 + 23 \log m_b^2/m^2_t + 25 \log m^2_c/m^2_b + 27\log 1\ \textrm{GeV}^2/m_c^2)
$$
Now, if one multiplies all of the quark masses by $1+{\delta v}/{v}$, one finds that
$$
\frac{1}{\alpha_{new}(1\ \GeV)^2} -
\frac{1}{\alpha_{old}(1\ \GeV)^2} = \frac{1}{12\pi}(21-27)*2\frac{\delta v}{v} = -\frac{1}{\pi}\frac{\delta v}{v} ,
$$
and plugging this into the standard one-loop formula for \qcd,
\begin{equation}
\Lambda^{2} \equiv\mu^{2}e^{-\frac{4\pi}{\beta_{0}\alpha_{s}\left(\mu^{2}\right)}},
\end{equation}
one finds that  
\begin{equation}
\Lambda_{new} = \mu\ \exp\frac{-6\pi}{27 \alpha_{new}} = \mu\ \exp\left( \frac{-6\pi}{27\alpha_{old}} + \frac{6}{27} \frac{\delta v}{v}\right) = \Lambda_{old} (1 + \frac{2}{9} \frac{\delta v}{v}),\end{equation}
which is close to the results in \cite{Agrawal:1997gf,Agrawal:1998xa}.  
Thus, our crude approximation yields \qcd\ scaling as $({\delta v}/{v})^{2/9}$

To be more precise, we integrate the 2 loop renormalization group equations.     The scale dependence of the strong coupling constant is controlled by the $\beta$-function which can be expressed as a perturbative series:
\begin{eqnarray}
\label{eq:alpha_s}
Q^{2}\frac{\partial}{\partial Q^{2}} \frac{\alpha_{s}}{4\pi} & =\beta\left(\alpha_{s}\right)=-\left(\frac{\alpha_{s}}{4\pi}\right)^{2}\sum_{n=0}\left(\frac{\alpha_{s}}{4\pi}\right)^{n}\beta_{n}.
\end{eqnarray}
The values of the first terms of the $\beta$-series are:
\begin{equation}
\beta_{0}=11-\frac{2}{3}n_{f},
\end{equation} 
at 1-loop level and
\begin{equation}
\beta_{1}=102-\frac{38}{3}n_{f},
\label{eq:beta_1}
\end{equation} 
at 2-loop level with $n_{f}$ the number of quark flavors active at the scale $Q^{2}$.

The exact analytical solution to Eq.~\eqref{eq:alpha_s} is known only to $\beta_{0}$ order, and thus we integrate numerically, defining $\Lambda_{QCD}$ to be the scale at which $\alpha_s$ diverges.   We find that  $\Lambda_{QCD}$ scales as $({\delta v}/{v})^{0.25}$, with the exponent varying from $0.245$ to $0.255$ over the entire possible range of \qcd.    Thus, we use $0.25$ as the exponent.


\end{document}